\documentclass[superscriptaddress,secnumarabic,amssymb,amsmath,12pt,nobibnotes,aps,prd,showkeys,nofootinbib]{revtex4}
\usepackage{amsmath}

\newtheorem{remark}{Remark}[section]
\usepackage{graphicx}
\usepackage{epsf}
\usepackage{bm}
\usepackage{amsmath}
\usepackage{amsfonts}
\usepackage{amssymb}
\usepackage{epstopdf}
\usepackage{natbib}
\usepackage{color}

\begin{document}
\tolerance=5000
\noindent
%--------------------------------------------------------------------
\title{Quintessential Inflation for Exponential Type Potentials: Scaling and Tracker Behavior}

\author{Llibert Arest\'e Sal\'o}
\email{l.arestesalo@qmul.ac.uk}
\affiliation{School of Mathematical Sciences, Queen Mary University of London, Mile End Road, London, E1 4NS, United Kingdom}

\author{Jaume Haro}
\email{jaime.haro@upc.edu}
\affiliation{Departament de Matem\`atiques, Universitat Polit\`ecnica de Catalunya, Diagonal 647, 08028 Barcelona, Spain}

\begin{abstract}
  
We will show that for exponential type potentials { of the form $V(\varphi)\sim e^{-\gamma\varphi^n/M_{pl}^n}$}, which are used to depict quintessential inflation,
the solutions whose initial conditions take place during the slow roll phase in order to describe correctly the inflationary period, 
do not belong { for large values of the parameter $n$}
 to the basin of attraction of the scaling solution -a solution of the scalar field equation whose energy density scale as the one of the fluid component of the universe during radiation or the matter domination period-, meaning that a late time mechanism to exit this behavior and depict correctly the current cosmic acceleration is not needed. { However, in such cases, namely $n$ large enough, these potentials cannot correctly depict the current cosmic acceleration. This is the reason why  the potential must be improved introducing another parameter -as the one in the well-known Peebles-Vilenkin quintessential inflation model, which depends on two parameters, one to describe inflation and the other one to correctly depict the present accelerated evolution-  able to deal with the late time acceleration of our universe.

}

\end{abstract}

\vspace{0.5cm}

%-------------------------------------------------------------------------
\maketitle
%------------------------------------------------------------------------
\thispagestyle{empty}
\tableofcontents

\section{Introduction}

Quintessence 
\cite{Caldwell:1997ii, rp, pr, barreiro,Carroll:1998zi,Chiba:1999wt,Sahni:1999qe} 
is a theory used to reproduce the current cosmic acceleration without the need of a cosmological constant.  In quintessence it has been shown that, for exponential
potentials $V(\varphi)\sim e^{-\gamma\varphi/M_{pl}}$ with
$\gamma>2$, there exists a solution whose energy density scales as the one of radiation \cite{copeland}. { Other successful quintessential inflation models have been found as well with potentials of similar behavior, such as the one in \cite{benisty}.} And it has recently been proved that for more general exponential potentials 
 $V(\varphi)\sim e^{-\gamma\varphi^n/M_{pl}^n}$ there also exists an approximate scaling solution \cite{hossain2, Geng:2017mic}. 
Such solution, termed as {\it scaling solution} (see \cite{liddle} for a detailed classification of the potentials that lead to scaling solutions), is important in order to deal with the coincidence problem because, due to the attractor behavior of the scaling solution, if the scalar field is at the beginning of radiation in the basin of attraction of this scaling solution, it evolves as a radiation fluid. Therefore, since in standard quintessence we have two fields, the inflaton (which vanishes after releasing its energy when it oscillates in the deep well of the potential) and the quintessence scalar field, one can assume initial conditions for this field which lead it to enter into the basin of attraction of the scaling solution.

\

However, so that the universe enters in  the late time accelerated phase, the quintessence field has to leave the scaling behavior, which could be done in several ways. Taking into account that for $0<\gamma<\sqrt{2}$ during the matter domination era there exists a {\it tracker solution} \cite{Steinhardt:1999nw, rp, copeland,UrenaLopez:2000aj} leading to an accelerating late time universe,
 one could add to the  potential the term
 $e^{-\gamma\varphi/M_{pl}}$, with $0<\gamma<\sqrt{2}$. 
 In this situation, it can be shown that the first term of the
 potential dominates during the radiation dominated era and the second term dominates during the matter dominated one \cite{barreiro}.  Alternatively, one could introduce a non-minimal coupling between the quintessence field and massive neutrinos, whose effect is to modify the potential in the matter domination era \cite{hossain2,Geng:2017mic}, but in that case, as we will see, the current cosmic acceleration is due to an effective cosmological constant.

\

On the contrary, in quintessential inflation \cite{pv,dimopoulos1,hossain1, 
hap1,
deHaro:2016hsh,deHaro:2016ftq,hap,deHaro:2017nui,AresteSalo:2017lkv,Haro:2015ljc,hap19} there is only one scalar field -the inflaton- driving the evolution of the universe by depicting both the early- and  late- acceleration of the universe. Due to the attractor behavior of inflation, the initial condition of the scalar field has to be taken to belong to the basin of attraction of the slow roll solution. Then, using a quintessential inflation model based on the exponential type potentials proposed in \cite{Geng:2017mic}, where the authors showed that there exists an approximately scaling solution, 
we  will show that { for large values of the parameter $n$}  at the beginning of the radiation era the scalar field is not in the basin of attraction of the scaling solution. 
In fact, the value of the effective Equation of State (EoS) parameter for the inflaton field is $1$ during the radiation epoch, that is, it does not scale as the relativistic plasma whose energy density dominates during this period.
As a consequence, in { such cases}  a mechanism to exit the scaling behavior is not needed. The only thing  needed to reproduce the evolution of the universe, { as Peebles and Vilenkin shown in its seminal paper \cite{pv}},  
is an inflationary potential leading to a spectral index ($n_s$) and a ratio of tensor-to-scalar perturbations ($r$) entering into the two dimensional marginalized joint confidence contour at $2\sigma$ confidence-level (CL) provided by Planck data  \cite{planck18,planck18a} combined with a quintessence  potential which is dominant at late times in order to correctly depict the current cosmic acceleration.

\

The paper is organized as follows. In Section \ref{sec-2}, we study the exponential type potentials introduced in \cite{hossain2}, we calculate the reheating temperature of the universe using the mechanism of {\it instant preheating} \cite{fkl0,fkl} (see \cite{ahmad} for more details) because the potential is very smooth and the gravitational particle production of neither light nor superheavy particles is effective for this kind of potentials \cite{ford, haro18,hashiba, Chung, Chung1}. With this reheating temperature we compute the evolution of the inflaton field during the kination regime \cite{Joyce} in order to obtain its initial conditions at the beginning of the radiation epoch. Finally, with this initial { data} we integrate numerically the dynamical system to show that { for $n$ sufficiently large (we have taken the value of $n=10$ to carry out the computations)}  the dynamics of the inflaton field is completely different to the one of the scaling solution.
Section \ref{sec3} is devoted to the study of a viable model of quintessential inflation 
whose potential is the combination of an exponential type potential -which stands for inflation- with a pure exponential potential which will reproduce the late time acceleration of the universe. To obtain numerically the value of the parameter on which the  model depends we use the current observation data such as the red-shift at the beginning of the matter-radiation equality, the current values of the Hubble rate and the ratio of the matter energy density to the critical one.
 Finally, in Section \ref{sec-summary} we present the conclusions of our work.

\

The units used throughout the paper are $\hbar=c=1$ and we denote  the reduced Planck's mass by 
$M_{pl}\equiv \frac{1}{\sqrt{8\pi G}}\cong 2.44\times 10^{18}$ GeV.

\section{A quintessential inflation model}
\label{sec-2}
In this work
we will consider the same  Exponential  Inflation-type potentials  studied, for the first time,  in  \cite{hossain2},
\begin{eqnarray}\label{exp}
V(\varphi)=\
V_0e^{-\lambda{\varphi}^n/{M_{pl}}^n},
\end{eqnarray}
where $\lambda$ is a dimensionless parameter and $n$ is an integer.

For this model the power spectrum of scalar perturbations, its spectral index and the ratio of tensor to scalar perturbations 
are given by (see for details of the calculations \cite{Geng:2017mic})
\begin{eqnarray}\label{power}
{\mathcal P}=\frac{V_0e^{-\lambda{\varphi}^n/{M_{pl}}^n}}{12\pi^2n^2\lambda^2M_{pl}^{6-2n}\varphi^{2n-2}}\sim 2\times 10^{-9},
\end{eqnarray}
\begin{eqnarray}\label{index}
n_s=1-n\lambda\left( \frac{\varphi}{M_{pl}}\right)^{n-2}\left( 2n-2+\lambda n\left( \frac{\varphi}{M_{pl}}\right)^{n} \right)
\end{eqnarray}
and
\begin{eqnarray}\label{r}
r=8n^2\lambda^2\left( \frac{\varphi}{M_{pl}}\right)^{2n-2}.
\end{eqnarray}

An important relation is obtained combining the equations (\ref{index}) and (\ref{r}),
\begin{eqnarray}\label{relation}
\lambda \left( \frac{\varphi}{M_{pl}}\right)^n=\frac{r(2n-2)}{n(8(1-n_s)-r) },
\end{eqnarray}
which leads to the formula for the power spectrum
\begin{eqnarray}
{\mathcal P}=\frac{2V_0e^{-\frac{r(2n-2)}{n(8(1-n_s)-r) }}}{3\pi^2M_{pl}^{4}r}\sim 2\times 10^{-9}
\end{eqnarray}
and, thus,
\begin{eqnarray}
V_0\sim 3\pi^2re^{\frac{r(2n-2)}{n(8(1-n_s)-r) }}\times 10^{-9} M_{pl}^4,
\end{eqnarray}
which for the viable values of $n_s=0.9649$ (the central value of the spectral index) and $r=0.02$ leads to 
\begin{eqnarray}\label{V0}
V_0\sim 6\pi^2(1.166)^{\frac{n-1}{n}}\times 10^{-11} M_{pl}^4.
\end{eqnarray}

\

It is important to realize that a way to find theoretically the possible values of the parameter $\lambda$ is to combine the equations (\ref{r}) and (\ref{relation}) to get
\begin{eqnarray}\label{lambda1}
\lambda=\left( \frac{n(8(1-n_s)-r)}{2r(n-1)} \right)^{n-1}\left(\frac{r}{8n^2}\right)^{n/2}.
\end{eqnarray}
And, using the theoretical values $r\leq 0.1$ and $n_s=0.9649\pm 0.0042$ (see for instance \cite{planck18,planck18a}), one can find the candidates of $\lambda$ at $2\sigma$ C.L. These values have to be checked for the joint contour in the plane $(n_s,r)$ at $2\sigma$ C.L., when the number of efolds is approximately between 60 and 75, which is what happens in quintessential inflation due to the kination phase \cite{leach,  deHaro:2016ftq} -the energy density of the scalar field is only kinetic \cite{Joyce, Spokoiny}- after the inflationary period.

\

For this kind of potentials, in order to compute the number of efolds we need to calculate the main slow-roll parameter 
\begin{eqnarray}
\epsilon=\frac{M_{pl}^2}{2}\left(\frac{V_{\varphi}}{V} \right)^2=\frac{\lambda^2n^2}{2}\left( \frac{\varphi}{M_{pl}} \right)^{2n-2},
\end{eqnarray}
whose value at the end of inflation is $\epsilon_{END}=1$, meaning that at the end of this epoch  the field reaches the value 
$\varphi_{END}=\left( \frac{2}{n^2\lambda^2}\right)^{\frac{1}{2n-2}} M_{pl}$. 

\

Then, the number of efolds is given by
\begin{eqnarray}\label{N}
N=\frac{1}{M_{pl}}\int_{\varphi}^{\varphi_{END}}\frac{1}{\sqrt{2\epsilon}}d\varphi=\frac{1}{n\lambda(n-2)}
\left[\left( \frac{\varphi}{M_{pl}} \right)^{2-n}-\left( \frac{2}{n^2\lambda^2}\right)^{\frac{2-n}{2n-2}} \right]
\end{eqnarray}
and, thus, combining the equations  (\ref{index}), (\ref{r}) and (\ref{N}) one obtains the spectral index and the tensor/scalar ratio as a function of the number of efolds and the parameter $\lambda$.

\

On the other hand, since
inflation ends at $\varphi_{END}=\left( \frac{2}{n^2\lambda^2}\right)^{\frac{1}{2n-2}} M_{pl}$, when the effective Equation of State (EoS) parameter is equal to $-1/3$,
meaning that $\dot{\varphi}_{END}^2=V(\varphi_{END})$,
the energy density at the end of inflation is
{
 \begin{eqnarray}\rho_{\varphi, END}=\frac{3}{2} V(\varphi_{END})=9\pi^2(1.166)^{\frac{n-1}{n}}e^{-\lambda \left( \frac{2}{n^2\lambda^2}\right)^{\frac{n}{2n-2}}}
 \times 10^{-11} M_{pl}^4
 \end{eqnarray}}
 and the corresponding value of the Hubble rate is given by
\begin{eqnarray}\label{HEND}
  H_{ END}=\sqrt{\frac{3}{10}}\pi(1.166)^{\frac{n-1}{2n}}e^{-\frac{\lambda}{2} \left( \frac{2}{n^2\lambda^2}\right)^{\frac{n}{2n-2}}}
 \times 10^{-5} M_{pl},
 \end{eqnarray}
which will constrain very much the values of the parameter $\lambda$ because in all viable inflationary models at the end of inflation the value of the Hubble rate is of the order of $10^{-6} M_{pl}$  \cite{linde}.
 In fact, when  (\ref{HEND}) is of the order  $10^{-6} M_{pl}$ one gets
 \begin{eqnarray}\label{lambda2}
 \lambda\sim \frac{(2/n^2)^{n/2}}{[\ln(30\pi^2 (1.166)^{(n-1)/n})]^{n-1}}.
 \end{eqnarray}

 \
 
  Then, to perform numerical calculations, throughout the paper we will use the values of
 $n=10$ and $r=0.02$ and, thus, for $\lambda=4.1\times 10^{-16}$ we have obtained approximately $67.4$ efolds, which is a viable value in quintessential inflation. {Note that the only constraint for $r$ is $r\leq 0.1$ and, therefore, we have been able to use the value which enables equations \eqref{lambda1} and \eqref{lambda2} to be compatible one to another, which turns out to be $r=0.02$ for $n=10$.}
 With these values, if the particles are created via instant preheating \cite{fkl0,fkl} -which seems the best mechanism due to the smoothness of the potential-  we have to obtain the Enhanced Symmetry Point (EPS), which is
 the value of the field at which its temporal derivative is maximum. In our case, taking initial conditions during the slow roll period (recall that the slow roll solution is an attractor, so the evolution of the inflation field is the same for all initial conditions in the basin of attraction of the slow roll solution) 
{we have obtained by integrating  numerically the dynamical system that approximately $\dot{\varphi}_{max}\cong 2.9\times 10^{-6} M_{pl}^2$  at $\varphi_{max}\cong 37 M_{pl}$.}
 
 \

 After this, we have to find out the moment when kination starts, which could be chosen at the moment when the effective EoS parameter is very close to $1$.
 Assuming for instance that kination starts at $w_{eff}\cong 0.99$, we have numerically obtained { $\varphi_{kin}\cong 47 M_{pl}$ and $\dot{\varphi}_{kin}\cong 5.6 \times 10^{-9} M_{pl}^2$, and hence $H_{kin}\cong 2.3\times 10^{-9} M_{pl}$.}
 
 \
 
On the other hand, since in instant preheating the effective mass of the quantum field $\chi$ field  coupled with the inflaton $\varphi$ is given by $g(\varphi-\varphi_{max})$, where $g$ is the dimensionless coupling constant,
 at the beginning of kination the energy density of the created superheavy particles   is given by 
 \begin{eqnarray}
 \rho_{\chi,kin}=g(\varphi_{kin}-\varphi_{max}) n_{\chi,max} \left( \frac{a_{max}}{a_{kin}} \right)^3=
 10g M_{pl} n_{\chi,max} \left( \frac{a_{max}}{a_{kin}} \right)^3,
 \end{eqnarray}
 where the density of produced particles is { \cite{fkl}}
 \begin{eqnarray}
 n_{\chi,max}=\frac{g^{3/2}\dot{\varphi}^{3/2}_{max}}{8\pi^3}\cong {1.99\times 10^{-11} g^{3/2}}M_{pl}^3
 \end{eqnarray}
 and 
 \begin{eqnarray}
 \left( \frac{a_{max}}{a_{kin}} \right)^3= e^{-3\int_{t_{max}}^{t_{kin}}H(t)dt}= e^{-3\int_{\varphi_{max}}^{\varphi_{kin}}\frac{H}{\dot{\varphi}}d\varphi}
 \cong {1.2\times 10^{-12}}
\end{eqnarray}
   has been calculated numerically.

\

Then, at the beginning of the kination  regime we have
\begin{eqnarray}
\rho_{\chi,kin}={ 2.39\times 10^{-22}} g^{5/2} M_{pl}^4 \quad \mbox{and} \quad \rho_{\varphi,kin}=  { 1.59\times 10^{-17}} M_{pl}^4
\end{eqnarray}
and, denoting by $\Gamma$ the decay rate -the superheavy particles must decay into lighter ones in order to obtain a relativistic plasma needed to match with the hot Friedmann universe-,  using that $\frac{H_{dec}}{H_{kin}}=\frac{\Gamma}{H_{kin}}=\left(\frac{a_{kin}}{a_{dec}} \right)^3$
and taking into account that the inflaton field $\varphi$ is nearly frozen during kination,
 we get
\begin{eqnarray}
\rho_{\varphi,dec}=3\Gamma^2M_{pl}^2\quad \mbox{and} \quad \rho_{\chi,dec}\cong {1.04\times 10^{-13}} g^{5/2}M_{pl}^3\Gamma.
\end{eqnarray}

\

In addition, in order to avoid a second inflationary phase we have to impose that the decay takes place before the end of the kination \cite{fkl} (see also \cite{haro19} for a detailed explanation), i.e., we have to assume
$\rho_{\chi,dec}\leq \rho_{\varphi,dec}$, which leads to the following constraint,
\begin{eqnarray}\label{bound}
\Gamma\geq { 3.47\times 10^{-14}} g^{5/2} M_{pl}.
\end{eqnarray}

\

Then, following for example Section II of \cite{haro19}, 
the reheating temperature is given by
\begin{eqnarray}
T_{rh}=\left(\frac{30}{\pi^2 g_{rh}} \right)^{1/4}\rho_{\chi,dec}^{1/4}\sqrt{\frac{\rho_{\chi,dec}}{\rho_{\varphi,dec}}}\cong
{ 4.34\times 10^{-11}}g^{15/8}\left(\frac{M_{pl}}{\Gamma} \right)^{1/4} M_{pl},
\end{eqnarray}
where $g_{rh}=106.75$ are the degrees of freedom for the Standard Model.

\

And, from the bound (\ref{bound}), we get the maximum value of the reheating temperature as
\begin{eqnarray}\label{temperaturebound}
T_{rh}\leq { 1.01\times 10^{-7}} g^{5/4}M_{pl}={ 2.45\times 10^{11}} g^{5/4} \mbox{ GeV}.\end{eqnarray}

\

On the other hand,  as we have already explained, the decay must be before the end of the kination phase, meaning  that $\Gamma\leq H_{kin}\cong { 2.3\times 10^{-9}} M_{pl}$, which leads to the lower bound 
\begin{eqnarray}
T_{rh}\geq { 6.27\times 10^{-9}} g^{15/8}M_{pl}={ 1.53 \times 10^{10}} g^{15/8} \mbox{ GeV}.\end{eqnarray}
Moreover,  to preserve the BBN success the reheating temperature has to be approximately constrained between $1$ MeV and $10^9$ GeV \cite{riotto}, so we get the bound
\begin{eqnarray}
{ 3.08}\times 10^{-12}\leq g\leq { 0.23}.
\end{eqnarray}

\

Finally, to fix the reheating temperature we choose the following compatible values of the parameters, $g=10^{-2}$ and $\Gamma=10^{-12} M_{pl}$, obtaining a reheating temperature of
\begin{eqnarray}
T_{rh}\cong { 7.72\times 10^{-12}} M_{pl}\cong { 1.88\times 10^7} \mbox{ GeV}.
\end{eqnarray}

\subsection{Dynamical evolution of the scalar field}

Next,  we want to calculate the value of the scalar field and its derivative at the reheating time.  
%Since the shape of the potential is very smooth, as we have already explained  the particle production must be  via instant
%preheating, 
% which means that the created particles have to decay into lighter ones before the end of the kination phase \cite{fkl0, fkl}.
Analytical calculations can be done disregarding the potential during kination because during this epoch the potential energy of the field is negligible. Then, since during kination one has $a\propto t^{1/3}\Longrightarrow H=\frac{1}{3t}$, using the Friedmann equation the dynamics in this regime will be
\begin{eqnarray}
\frac{\dot{\varphi}^2}{2}=\frac{M_{pl}^2}{3t^2}\Longrightarrow \dot{\varphi}=\sqrt{\frac{2}{3}}\frac{M_{pl}}{t}\Longrightarrow 
\varphi(t)=\varphi_{kin}+\sqrt{\frac{2}{3}}M_{pl}\ln \left( \frac{t}{t_{kin}} \right).\end{eqnarray}

Thus, at the reheating time, i.e., at the beginning of the radiation phase, one has 
\begin{eqnarray}
\varphi_{rh}=\varphi_{kin}+\sqrt{\frac{2}{3}}M_{pl}\ln\left( \frac{H_{kin}}{H_{rh}} \right).
\end{eqnarray}
And, using that at the reheating time (i.e., when the energy density of the scalar field and the one of the relativistic plasma coincide) 
the Hubble rate is given by $H_{rh}^2=\frac{2\rho_{rh}}{3M_{pl}^2}$, one gets 
\begin{eqnarray}
\varphi_{rh}=\varphi_{kin}+\sqrt{\frac{2}{3}}M_{pl}\ln\left( \frac{ H_{kin}}{\sqrt{\frac{\pi^2g_{rh}}{45}} \frac{T_{rh}^2}{M_{pl}}}\right)
\qquad \mbox{and}  \qquad  { \dot{\varphi}_{rh}=\sqrt{\frac{\pi^2g_{rh}}{15}} T_{rh}^2},\end{eqnarray} 
where we have used that  the energy density and the temperature are related via  the formula
 $\rho_{rh}=\frac{\pi^2}{30}g_{rh}T_{rh}^4$, where the number of degrees of freedom for the Standard Model is $g_{rh}=106.75$  \cite{rg}.

{
As we have already commented, we will take as the reheating temperature $T_{rh}\cong 2\times 10^{-7}$ GeV. Then, at the beginning of the radiation era we will have
}
 \begin{eqnarray} \label{xrh}
 \varphi_{rh}\cong {  71.3}M_{pl} \qquad {\dot{\varphi}_{rh}\cong  { 4.99\times 10^{-22}} M_{pl}^2}.
 \end{eqnarray}

 To calculate the value of the field and its derivative at the matter-radiation equality, namely $\varphi_{eq}$ and $\dot{\varphi}_{eq}$,
 we continue assuming that the potential is negligible (this situation has to be verified numerically integrating the full dynamical system, which we will show in the next subsection), i.e., we are assuming that
 the effective EoS parameter of the field $w_{\varphi}=(\dot{\varphi}^2/2 -V)/(\dot{\varphi}^2/2 +V)=1$, namely that the field never reaches the basin of attraction of
 the scaling solution (which will be proved numerically in the next subsection), which should have during the radiation epoch an effective EoS parameter equal to $1/3$ because it scales as a radiation fluid.
 
  Now, we consider the central values obtained in \cite{planck} (see the second column in Table $4$) of  the red-shift at the matter-radiation equality $z_{eq}=3365$,
the present value of the ratio of the matter energy density to the critical one $\Omega_{m,0}=0.308$, and $H_0=67.81\; \mbox{Km/sec/Mpc}=5.94\times 10^{-61} M_{pl}$.
Then, the present value of the matter energy density is $\rho_{m,0}=3H_0^2M_{pl}^2\Omega_{m,0}=3.26\times 10^{-121} M_{pl}^4$ and at the matter-radiation equality we will 
have $\rho_{eq}=2\rho_{m,0}(1+z_{eq})^3=2.48\times 10^{-110} M_{pl}^4=8.8\times 10^{-1} \mbox{eV}^4$. 
Now,  using the relation at the matter-radiation equality $\rho_{eq}=\frac{\pi^2}{15}g_{eq}T_{eq}^4$ with $g_{eq}=3.36$ (see \cite{rg}), we  get 
$T_{eq}=3.25\times 10^{-28} M_{pl}=7.81\times 10^{-10}$ GeV. Thus, 
solving the dynamical equation $\ddot{\varphi}+\frac{3}{2t}\dot{\varphi}=0$, 
one  obtains
\begin{eqnarray}\label{xeq}
 \varphi_{eq}=\varphi_{rh}+2\sqrt{\frac{2}{3}}M_{pl}\left(1-\sqrt{\frac{2H_{eq}}{3H_{rh}}}\right)
 =\varphi_{rh}+2\sqrt{\frac{2}{3}}M_{pl}\left(1-\sqrt{\frac{2}{3}}\left( \frac{g_{eq}}{g_{rh}} \right)^{1/4}\frac{T_{eq}}{T_{rh}}\right) \nonumber \\
  \cong \varphi_{rh}+2\sqrt{\frac{2}{3}}M_{pl} 
  \cong {  72.9 }M_{pl}.
  \end{eqnarray}
 \begin{eqnarray}\label{yeq}
\dot{\varphi}_{eq}=\dot{\varphi}_{rh}\frac{t_{rh}}{t_{eq}}\sqrt{\frac{t_{rh}}{t_{eq}}}=\frac{4}{3}M_{pl}H_{eq}\sqrt{\frac{H_{eq}}{H_{rh}}}
=\frac{4\pi}{9}\sqrt{\frac{g_{eq}}{5}}\left(\frac{g_{eq}}{g_{rh}} \right)^{1/4}\frac{T_{eq}^3}{T_{rh}}\cong  { 1.28\times 10^{-17} \mbox{ eV}^2},
\end{eqnarray} 
where once again we have used that $T_{rh}\cong 2\times 10^7$ GeV.

\subsection{Numerical simulation during radiation}

To show numerically that  during radiation the scalar field is not in the basin of attraction of the scaling solution, 
first of all we calculate the value of the red-shift at the beginning of the radiation epoch
\begin{eqnarray}
1+z_{rh}=\frac{a_0}{a_{rh}}=\frac{a_0}{a_{eq}}\frac{a_{eq}}{a_{rh}}=
(1+z_{eq})\left(\frac{\rho_{r,rh}}{\rho_{r,eq}}  \right)^{1/4}= (1+z_{eq})\left(\frac{g_{rh}}{g_{eq}}  \right)^{1/4}\frac{T_{rh}}{T_{eq}},
\end{eqnarray}
where  we have used that $\rho_{r,eq}=\rho_{r,rh}\left(\frac{a_{rh}}{a_{eq}} \right)^4$.
Then, for the reheating temperature $T_{rh}=2\times 10^7$ GeV, we get  $z_{rh}=-1+{ 1.90\times 10^{20}}$.

\

Moreover, at the beginning of radiation the energy density of the matter will be
\begin{eqnarray}
\rho_{m,rh}=\rho_{m,eq}\left( \frac{a_{eq}}{a_{rh}} \right)^3=\rho_{m,eq}\left(\frac{\rho_{r,rh}}{\rho_{r,eq}}  \right)^{3/4}
=\frac{\pi^2}{30}g_{eq}\left( \frac{g_{rh}}{g_{eq}} \right)^{3/4}T^3_{rh}T_{eq}\nonumber \\ \cong { 7.7\times 10^{13}} \mbox{ GeV}^4,
\end{eqnarray}
where we have used that $\rho_{m,eq}=\rho_{r,eq}=\rho_{eq}/2$.

\

In this way,  
the dynamical equations after the beginning of the radiation can be easily obtained using as a time variable
$N\equiv -\ln(1+z)=\ln\left( \frac{a}{a_0}\right)$. Recasting the  energy density of radiation and matter respectively as functions of $N$, we get
\begin{eqnarray}
\rho_{r}(a)={\rho_{r,rh}}\left(\frac{a_{rh}}{a}  \right)^4\Longrightarrow \rho_{r}(N)= {\rho_{r,rh}}e^{4(N_{rh}-N)} 
\end{eqnarray}
and
\begin{eqnarray}
\rho_{m}(a)={\rho_{m,rh}}\left(\frac{a_{rh}}{a}  \right)^3\Longrightarrow \rho_{m}(N)={\rho_{m,rh}}e^{3(N_{rh}-N)},
\end{eqnarray}
where  
$N_{rh}$ denotes the value of the time $N$ at the beginning of radiation and,  as we have already obtained, $\rho_{m,rh}\cong { 7.7\times 10^{13}} \mbox{ GeV}^4$ and 
$\rho_{r,rh}\cong { 4.4\times 10^{30}} \mbox{ GeV}^4$.

\

To obtain the dynamical system for this scalar field model, we will
introduce the  dimensionless variables
 \begin{eqnarray}
 x=\frac{\varphi}{M_{pl}} \qquad \mbox{and} \qquad y=\frac{\dot{\varphi}}{K M_{pl}},
 \end{eqnarray}
 where $K$ is a parameter that we will choose accurately in order to ease the numerical calculations. So,  taking into account  the conservation equation $\ddot{\varphi}+3H\dot{\varphi}+V_{\varphi}=0$, one arrives at the following   dynamical system,
 \begin{eqnarray}\label{system}
 \left\{ \begin{array}{ccc}
 x^\prime & =& \frac{y}{\bar H}~,\\
 y^\prime &=& -3y-\frac{\bar{V}_x}{ \bar{H}}~,\end{array}\right.
 \end{eqnarray}
 where the prime is the derivative with respect to $N$, $\bar{H}=\frac{H}{K}$   and $\bar{V}=\frac{V}{K^2M_{pl}^2}$.  It is not difficult to see that  one can write  
 \begin{eqnarray}
 \bar{H}=\frac{1}{\sqrt{3}}\sqrt{ \frac{y^2}{2}+\bar{V}(x)+ \bar{\rho}_{r}(N)+\bar{\rho}_{m}(N) }~,
 \end{eqnarray}
where we have defined the dimensionless energy densities as
 $\bar{\rho}_{r}=\frac{\rho_{r}}{K^2M_{pl}^2}$ and 
 $\bar{\rho}_{m}=\frac{\rho_{m}}{K^2M_{pl}^2}$.

\

Next, to integrate from the beginning of radiation up to the matter-radiation equality, i.e.,   from $N_{rh}\cong -50.57$ to $N_{eq}\cong -8.121$,
 we will choose $KM_{pl}= 10^{-17} \mbox{GeV}^2$, yielding 
{\begin{eqnarray}
\bar{\rho}_{r}(N)= { 4.4\times10^{64}}e^{4(N_{rh}-N)}, \qquad \bar{\rho}_{m}(N)= { 7.7 \times10^{47}}e^{3(N_{rh}-N)}
\end{eqnarray}}
and
{\begin{eqnarray}
\bar{V}(x) = \frac{ V_0\times 10^{34}}{\mbox{GeV}^4} e^{-\lambda x^n}.
\end{eqnarray}}

Finally, we use the initial conditions for the field as { $x_{rh}=71.3$ and $y_{rh}=2.97 \times 10^{32}$ (they have been obtained in the equation \eqref{xrh}. Note that we could also have used the initial conditions in $N_{rh}$ computed in \eqref{xeq} and \eqref{yeq}, but the assumptions applied in this calculus might not be valid at all when including a new exponential term as done in next section.}
By integrating numerically the dynamical system, we conclude that, { for values of the parameter $n$ greater enough,  the value of  the EoS parameter $w_{\varphi}$ remains $1$ during radiation, namely between $N_{rh}$ and $N_{eq}$, thus proving that the inflaton field does not belong to the basin of attraction of the scaling solution in these cases.}

\

\section{A viable model}
\label{sec3}

To depict the late time acceleration, we have to modify the original potential because it cannot explain the current observational data
{(for the potential (1), at the present time  the    density parameter   $\Omega_{\varphi}=\frac{\rho_{\varphi}}{3H^2 M_{pl}^2}$ is far from  its observational value, namely $0.7$)}. For this reason {and following the spirit of the Peebles-Vilenkin model \cite{pv}, to match with the current observational data we have to introduce a new parameter  $M$  with units of mass which must be calculated numerically. In our case, we will consider the following modification of the potential (1) by introducing a new exponential term containing the parameter $M$,}
\begin{eqnarray}\label{viable}
V(\varphi)=\
V_0e^{-\lambda{\varphi}^n/{M_{pl}}^n}+ M^4e^{-\gamma \varphi/M_{pl}},
\end{eqnarray}
with  $0<\gamma<\sqrt{2}$ in order to obtain that at late times the solution is in the basin of attraction of the tracker solution
\cite{Steinhardt:1999nw, UrenaLopez:2000aj}, which evolves as a fluid with effective EoS  parameter $w_{eff}=\frac{\gamma^2}{3}-1$ (see \cite{hap19a} for
a  detailed deduction of the tracker solution).

\

{
\begin{remark}
An important remark is in order. Our potential slightly differs from the one used in \cite{hossain2, Geng:2017mic}
\begin{eqnarray}
V(\varphi)=\
V_0e^{-\lambda{\varphi}^n/{M_{pl}}^n}+ (\bar{\rho}_{\nu}-3\bar{p}_{\nu})e^{\beta \varphi/M_{pl}},
\end{eqnarray}
where $\rho_{\nu}=\bar{\rho}_{\nu}e^{\beta \varphi/M_{pl}}$  is the  neutrino  energy density with constant bare mass $m_{\nu}$ and 
 $\beta>0$ is the non-minimal coupling of neutrinos with the inflaton field (see for details \cite{ Geng:2017mic}), which has a minimum where inflation ends its evolution, thus acting as an effective cosmological constant which stands for the current cosmic acceleration. In contrast, our potential does not have a minimum and the scalar field continues rolling down the potential as happens in quintessential inflation.
 
 \
 
 On the other hand, in our case we can also justify our choice   considering that the inflaton field is non-minimally coupled with neutrinos but with a negative coupling constant, namely $\beta=-\gamma<0$. Therefore, $M^4=\bar{\rho}_{\nu}-3\bar{p}_{\nu}$ and the effective mass of neutrinos is given by  
 $m_{\nu,eff}(\varphi)= m_{\nu}e^{-\gamma \varphi/M_{pl}}$, which tends to zero for large values of the field. Hence, neutrinos become relativistic, contrary to what happens when $\beta$ is positive, where the neutrinos acquire a heavy mass becoming non-relativistic particles.
 Finally, to match with the current observational data, 
 $M^4=\bar{\rho}_{\nu}-3\bar{p}_{\nu}$ must be very small compared to the Planck's energy density.
 In fact,  for $n=10$ and $\gamma=1$ we have numerically obtained $M^4\sim 10^{-4} \mbox{MeV}^4$.
 \end{remark}

Now we have to solve the dynamical system (\ref{system}) with initial conditions at the beginning of the radiation era. Choosing  $K=H_0$ (the present value of the Hubble rate), the initial conditions become $x_{rh}=71.3$ and $y_{rh}=8.57 \times 10^{38}$. With this choice of the constant $K$, note that
 in order obtain of the parameter $M$ one has to simply impose the value of $\bar{H}=H(N)/H_0$ to be $1$ at $N=0$. 

}

\subsection{Numerical calculations}

{Choosing for instance $\gamma=1$},
integrating the dynamical system and imposing that $\bar{H}(0)=1$, we have obtained, { in the case $n=10$}, $M=1.94 \times 10^{-1}$ MeV. Thus, looking at formula (\ref{V0}), we realize that
during the slow roll phase ($\varphi\sim M_{pl}$) the second term of the potential (\ref{viable}) is sub-leading, that is, the responsible for inflation is the first term. On the contrary, for large values of the inflaton field $\varphi$ ($\varphi\sim 70 M_{pl}$), the second term of the potential is dominant, meaning that  it is the responsible for quintessence.

\

On the other hand, in Figure $1$ we show the evolution of the $\Omega$'s { for radiation, matter and the scalar field}. One can see that the energy density of the scalar field is dominant { at the present time and future.  Moreover, in} Figure $2$ one can deduce that the universe is accelerating because the effective EoS parameter is at the present time and future  less than $-1/3$.

\begin{figure}
\begin{center}
\includegraphics[scale=0.7]{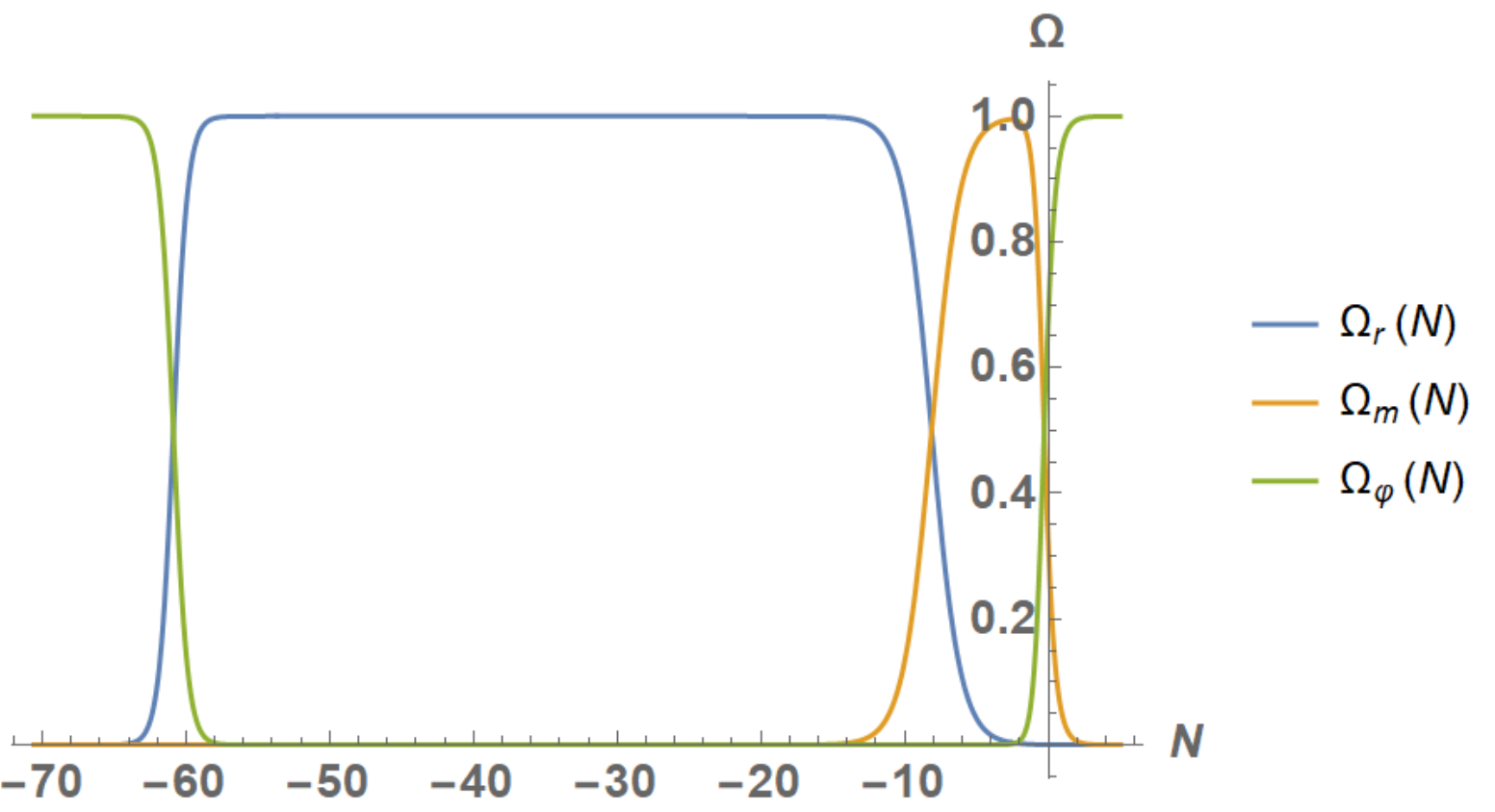}
\end{center}
\caption{
%{\em Left graph:} The reduced densities $\bar{\rho}_m$ (blue curve), $\bar{\rho}_r$ (red curve), and $\bar{\rho}_{\varphi}$ (green curve) in log scale, from reheating to matter-radiation equality. 
%{\em Right graph:} 
{ The density parameters $\Omega_m=\frac{\rho_m}{3H^2 M_{pl}^2}$ (orange curve), $\Omega_r=\frac{\rho_r}{3H^2 M_{pl}^2}$ (blue curve) and  
$\Omega_{\varphi}=\frac{\rho_{\varphi}}{3H^2 M_{pl}^2}$ (green curve), from kination to future times. To perform numerical calculations we have taken $n=10$ and $\gamma=1$.}
}
\label{fig:PV_rh_eq}
\end{figure}

\begin{figure}
\begin{center}
\includegraphics[scale=0.7]{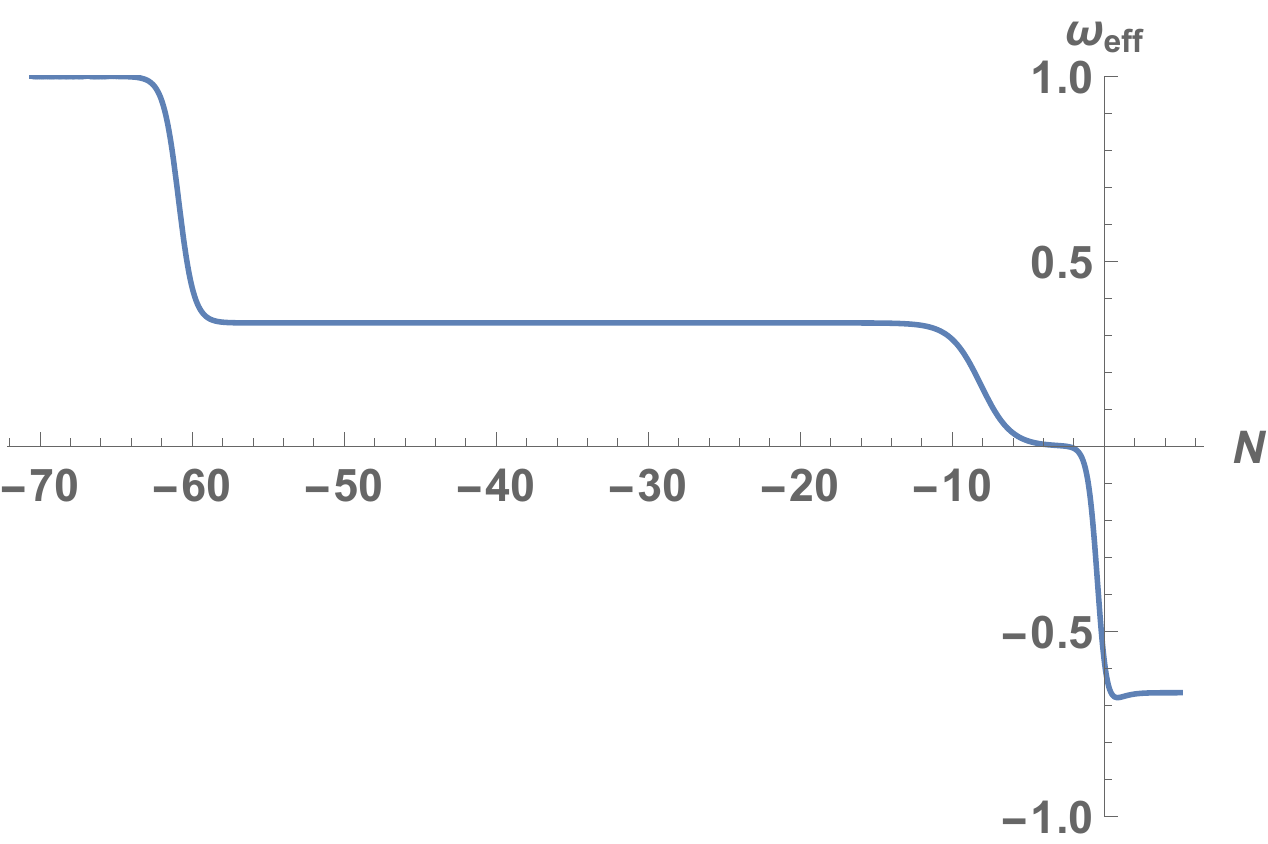}
\end{center}
\caption{
{The effective Equation of State parameter $w_{eff}$, from kination to future times, for $n=10$ and $\gamma=1$. As one can see in the picture, after kination the universe enters in a large period of time
where radiation dominate. Then, after the matter-radiation equality, the universe becomes matter-dominated and,  finally,  near the present, it  enters in a new accelerated phase
where $w_{eff}$ approaches $\frac{1}{3}-1=-\frac{2}{3}$, that is, it has the same effective EoS parameter as the tracker solution, meaning that the solution is in the basin of attraction of the tracker one.}}
\label{fig:PV_rh_eq}
\end{figure}

\

\section{Concluding remarks}
\label{sec-summary}

 We have shown that in quintessential inflation with exponential type potentials $V\sim e^{-\lambda \varphi^n/M_{pl}^n}$, { for large values of the parameter $n$} the solutions obtained from initial conditions during the slow roll regime do not enter in the basin of attraction of the scaling solution.
 
 In addition, we have seen that these potentials only depict the inflationary period. So, to obtain the current cosmic acceleration and describing all the evolution of the universe, we need to combine them with a quintessential potential. In our work, we have chosen as a quintessential potential an exponential potential of the type $e^{-\gamma\varphi/M_{pl}}$ with $0<\gamma<\sqrt{2}$ in order that at late times the solution is in the basin of attraction of the tracker solution, thus depicting a late time accelerated universe.

\vspace{1cm}

{\it Acknowledgments.}   
This investigation has been supported by MINECO (Spain) grant  MTM2017-84214-C2-1-P, and  in part by the Catalan Government 2017-SGR-247.

\end{document}